\documentclass[conference]{IEEEtran}
\IEEEoverridecommandlockouts

\usepackage{cite}
\usepackage[colorlinks]{hyperref}
\usepackage[pagebackref=true,breaklinks=true,colorlinks,bookmarks=false]{}
\usepackage{multirow}
\usepackage{amsmath,amssymb,amsfonts}
\usepackage{algorithmic}
\usepackage{graphicx}
\usepackage{textcomp}
\usepackage{xcolor}
\usepackage{url}
\usepackage{balance}
\def\BibTeX{{\rm B\kern-.05em{\sc i\kern-.025em b}\kern-.08em
    T\kern-.1667em\lower.7ex\hbox{E}\kern-.125emX}}

\begin{document}
\title{Relating CNN-Transformer Fusion Network for Change Detection
}

\author{
\IEEEauthorblockN{Yuhao Gao,
Gensheng Pei, Mengmeng Sheng,
Zeren Sun,
Tao Chen
and Yazhou Yao} 
\IEEEauthorblockA{School of Computer Science and Engineering, Nanjing University of Science and Technology, Nanjing, China\\ \{yuhao\_0, peigsh, shengmengmeng, zerens, taochen, yazhou.yao\}@njust.edu.cn } 
}

\maketitle

\begin{abstract}
	
While deep learning, particularly convolutional neural networks (CNNs), has revolutionized remote sensing (RS) change detection (CD), existing approaches often miss crucial features due to neglecting global context and incomplete change learning. Additionally, transformer networks struggle with low-level details. RCTNet addresses these limitations by introducing \textbf{(1)} an early fusion backbone to exploit both spatial and temporal features early on, \textbf{(2)} a Cross-Stage Aggregation (CSA) module for enhanced temporal representation, \textbf{(3)} a Multi-Scale Feature Fusion (MSF) module for enriched feature extraction in the decoder, and \textbf{(4)} an Efficient Self-deciphering Attention (ESA) module utilizing transformers to capture global information and fine-grained details for accurate change detection. Extensive experiments demonstrate RCTNet's clear superiority over traditional RS image CD methods, showing significant improvement and an optimal balance between accuracy and computational cost. Our source codes and pre-trained models are available at: \textcolor{red}{\url{https://github.com/NUST-Machine-Intelligence-Laboratory/RCTNet}}.
 
\end{abstract}


\section{Introduction} \label{sec:intro}

\begin{figure}[t]
\centering 
\includegraphics[width=0.90\linewidth]{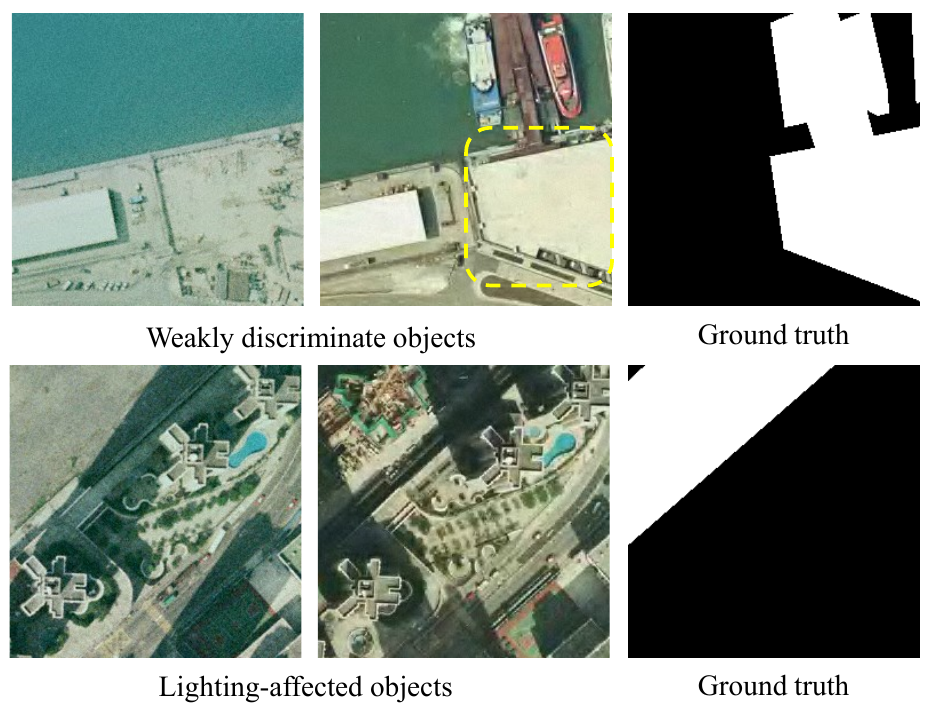}
\caption{The illustration of challenging scenarios, \textit{e.g.}, weakly discriminate objects and lighting-affected objects. The characteristics of the labeled building presented in the first example are similar to that of the ground. The second example shows shadow interference in changed regions.}
\vspace{-0.3cm} 
\label{img:moti}
\end{figure}

Change detection (CD) in remote sensing (RS) images stands as a critical technique for identifying semantic changes like building development or land cover modifications across geographically identical areas captured at different times. Its applications are extensive applications support diverse fields like damage assessment, urban planning, and natural disaster monitoring. RS images often encompass complex backgrounds and are susceptible to lighting variations. 
Fig.~\ref{img:moti} illustrates the challenges faced by the existing remote sensing change detection.
To address these challenges, researchers have employed a variety of techniques. Early CD methods relied on algorithms like Change Vector Analysis (CVA), Support Vector Machines (SVM), Kauth-Thomas (KT), and Principal Component Analysis (PCA), which demanded complex feature engineering and faced limitations in generalizability.

\begin{figure*}[t]
	\centering
	\includegraphics[width=1.0\textwidth]{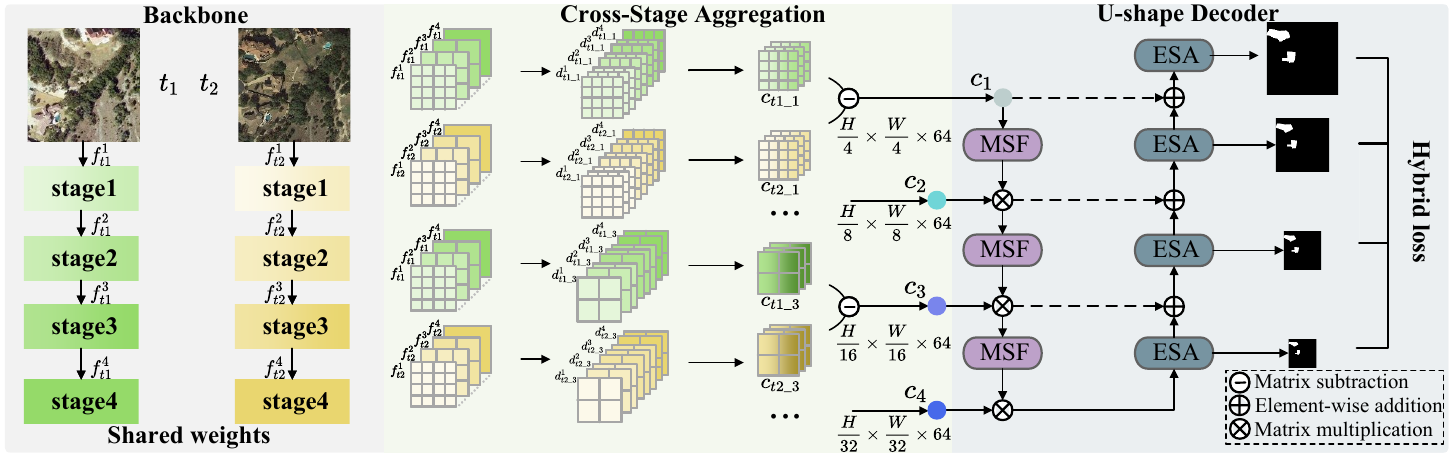}
	\caption{An illustration of our RCTNet. Our proposed architecture comprises three key modules: a shared-weight backbone for temporal feature extraction, a Cross-Stage Aggregation (CSA, \S\ref{sec:csa}) module for enhanced representation at each stage, and a U-shape decoder utilizing Multi-Scale Fusion (MSF, \S\ref{sec:pa_msf}) and Efficient Self-deciphering Attention (ESA, \S\ref{sec:pa_msf}) for robust decoding. RegNet extracts features from a registered image pair, and the CSA module enriches each stage's output. Finally, the U-shape decoder fuses multi-scale features through MSF and leverages ESA for accurate predictions.}
	\label{img:float}
\end{figure*}

Modern solutions have embraced the power of Convolutional Neural Networks (CNNs) for RS image CD tasks. Daudt \textit{et al.}~\cite{daudt2018fully} pioneer U-Net-inspired Siamese Fully Convolutional Networks (FCNs) for CD, establishing a foundational framework with its investigation of skip connection strategies. Hafner \textit{et al.}~\cite{hafner2021sentinel} utilize a dual-stream U-Net for data fusion from Sentinel-1 and Sentinel-2 images, while DASNet~\cite{chen2020dasnet} incorporates the extended attention mechanism for enhancing building change detection.
However, these approaches with expanded receptive fields often encounter a trade-off between global representation and computational complexity inherent to CNNs. This spurs explorations into Transformers~\cite{dosovitskiy2020image}, renowned for their exceptional ability to model global information relationships. The technologyy, previously reserved for natural language processing, has seen rapid advancements in vision tasks, \textit{e.g.}, image classification~\cite{yao2020towards,yao2021jo,sun2021webly,yao2017exploiting,sheng2024adaptive,sun2022pnp,liu2021exploiting}, object detection~\cite{yao2023automated,cai2024poly,zhang2023difference}, and segmentation \cite{pei2022hierarchical,liu2023fecanet,yao2021non,pei2023hierarchical,chen2023multi,chen2024spatial,pei2024videomac,10105896,chen2022saliency,chen2021semantically}. This led to the integration of Transformers in bitemporal RS image CD, with SwinSUNet~\cite{zhang2022swinsunet} attempting the pure transformer but encountering computational challenges. Chen \textit{et al.}~\cite{chen2021remote} propose the Bitemporal Image Transformer (BIT) method, combining CNN and transformers for feature extraction, yet its single-scale nature limits performance in subtle change regions. Bandara \textit{et al.}~\cite{bandara2022transformer} address this with a hierarchical transformer encoder coupled with a lightweight MLP decoder, effectively exploiting multi-layer features but with less efficient spatiotemporal detail correlation.

Motivated by the complementary nature of high-level semantic information and low-level detailed features crucial for change detection, we propose RCTNet, a hybrid network integrating the strengths of CNNs and Transformers. At its core lies the Cross-Stage Aggregation (CSA) module, which fuses features from different backbone stages. This fusion enriches both semantic information and fine-grained details of objects, empowering the model to capture the broader context and the intricate nuances within the image. An element-wise subtraction followed by an absolute value operation then computes temporal difference features, highlighting changes between the bitemporal images. 
These aggregated and enriched features are subsequently fed into the lightweight U-shape decoder. This decoder contains two core modules: Multi-Scale Feature Fusion (MSF) and Efficient Self-deciphering Attention (ESA). MSF further enhances the expressive capacity of the features by extracting information at multiple scales. At the same time, ESA incorporates global semantic relationships into each decoder layer, allowing the model to capture complex object changes effectively and ultimately boosting accuracy. Our contributions can be summarized as follows:

(1) We introduce a new CSA module that seamlessly integrates features from diverse stages of the backbone network. This fusion enriches both semantic information and fine-grained details across high-level and low-level feature maps, empowering the model to capture both context and intricate changes.

(2) We develop a lightweight U-shape decoder comprising two core modules: MSF and ESA. MSF improves semantic relationships between output features across different levels via multi-scale feature extraction, while ESA focuses on enhancing feature accuracy by computing global semantic relationships at each layer.

(3) We rigorously evaluate RCTNet on three representative CD datasets. Compared to existing state-of-the-art models, RCTNet achieves superior or highly competitive performance across all benchmark datasets, demonstrating its exceptional effectiveness in change detection tasks.

\section{Methodology}

\subsection{Overview} \label{sec:overview}
Fig.~\ref{img:float} depicts the overall architecture of RCTNet. The input bitemporal images, $t_1$ and $t_2$, of size $H \times W \times 3$ are fed into a shared-weight Siamese network to extract features. RCTNet downsamples the images through four stages, each containing a convolutional layer with stride 2. This progressive downsampling enables the network to learn semantic features at different scales, enhancing training stability. Each stage also utilizes max pooling for downsampling and feature preservation, facilitating the extraction of higher-level features.

RCTNet employs the lightweight RegNet~\cite{radosavovic2020designing} for feature extraction. 
Our proposed Cross-Stage Aggregation (CSA) module fuses features from different backbone stages, enriching semantic information and fine-grained details across feature maps. A lightweight U-shape decoder with two core modules further refines the extracted features. Multi-Scale Feature Fusion (MSF) enhances semantic relationships by extracting multi-scale features, while Efficient Self-deciphering Attention (ESA) incorporates global semantic connections into each decoder layer, ultimately boosting model accuracy.

\subsection{Cross-Stage Aggregation} \label{sec:csa}

\begin{figure}[t]
\centering 
\includegraphics[width=1\linewidth]{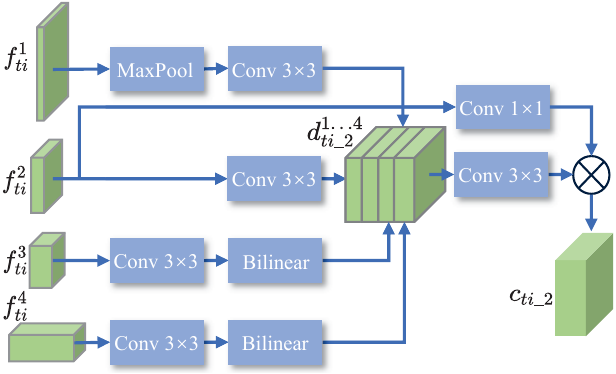}
\caption{Illustration of cross-stage aggregation (CSA).} \label{img:intro}
\label{img:csf}
\end{figure}

The Cross-Stage Aggregation (CSA) module comprises four parallel branches, each processing features extracted from different stages of the backbone for the corresponding bitemporal image. Focusing on the second branch (see Fig.~\ref{img:csf}), we illustrate how it transforms feature maps $f_{ti}^1, f_{ti}^2, f_{ti}^3, f_{ti}^4$ (extracted from stages $i=1,2$ of image $t_i$) into the output feature map $c_{ti\_2}$.
To align spatial resolutions, $f_{ti}^1$ undergoes max pooling and a $3\times3$ convolutional layer with channel adjustment.
Similarly, $f_{ti}^2$ receives a $3\times3$ convolutional layer for channel matching.
Features $f_{ti}^3$ and $f_{ti}^4$ are downsampled using bilinear upsampling and $3\times3$ convolutional layers for channel reduction, aligning them with the intermediate resolution.
Finally, all aligned feature maps are concatenated and processed by a final $3\times3$ convolutional layer with batch normalization and ReLU activation, generating the output $c_{ti\_2}$.

The CSA module carefully balances feature quantity and detail by allocating the following channel numbers to the four branches: 32, 64, 32, and 16. This configuration retains high-resolution features from shallower layers (32 channels) while incorporating semantic information from deeper layers (64 channels). In effect, the module leverages both details and context to enrich the aggregated representation. The complete feature transformation process can be summarized as:
\begin{eqnarray}
\begin{aligned}
d_{ti\_2}^1 & =\operatorname{Conv}_{3 \times 3}\left(\operatorname{Maxpool}\left(f_{ti}^1\right)\right)\in \mathbb{R}^{H/8 \times W/8 \times 32}, \\
d_{ti\_2}^2 & =\operatorname{Conv}_{3 \times 3}\left(f_{ti}^2\right)\in \mathbb{R}^{H/8 \times W/8 \times 64}, \\
d_{ti\_2}^3 & =\operatorname{Up}\left(\operatorname{Conv}_{3 \times 3}\left(f_{ti}^3\right)\right)\in \mathbb{R}^{H/8 \times W/8 \times 32}, \\
d_{ti\_2}^4 & =\operatorname{Up}\left(\operatorname{Up}\left(\operatorname{Conv}_{3 \times 3}\left(f_{ti}^4\right)\right)\right)\in \mathbb{R}^{H/8 \times W/8 \times 16}.
\end{aligned}
\end{eqnarray}
Here, $\operatorname{Maxpool}\left(\cdot\right)$ represents the max-pooling operation, and $\operatorname{Up}\left(\cdot\right)$ denotes bilinear upsampling. Each $d_{ti\_2}^1$ is a feature map extracted from the $i$-th stage of the backbone network for image $t_i$. These features are then combined through concatenation to capture richer spatial context information.

Following the feature transformation in the CSA branches, we further refine both $f_{ti}^2$ and the concatenated features. A $1 \times 1$ convolutional layer projects $f_{ti}^2$ to a consistent embedding dimension, while a $3 \times 3$ convolutional layer transforms the concatenated features. These processed representations then undergo a matrix multiplication operation, enabling the fusion of global and local information captured by the different components, formalized as:
\begin{equation}
c_{ti\_2} = {f_{ti}^2 }'\otimes\operatorname{Conv}_{3 \times 3}\left(\operatorname{Cat}\left(d_{ti\_2}^1,d_{ti\_2}^2,d_{ti\_2}^3,d_{ti\_2}^4\right)\right),
\end{equation}
where $\otimes$ indicates matrix multiplication, and $\operatorname{Cat}$ denotes feature concatenation. $c_{ti\_2}$ is the final output feature map, and ${f_{ti}^2 }'$ is the transformed $f_{ti}^2$. Similar operations are performed for the other CSA branches, enabling information integration across different spatial scales and semantic levels.

\begin{figure}[t]
\centering 
\includegraphics[width=0.89\linewidth]{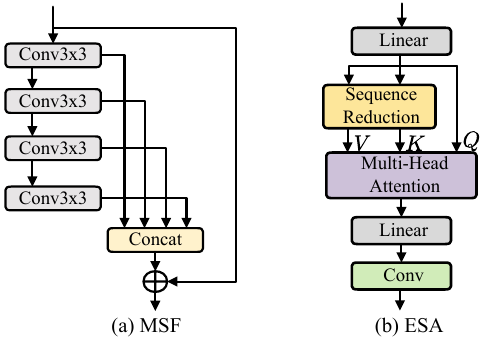}
\caption{Illustration of the two modules in U-shape decoder.}
\label{img:U}
\end{figure}

\begin{table*}[t]
    \begin{center}
        \caption{Quantitative comparisons on the WHU-CD~\cite{ji2018fully}, LEVIR-CD~\cite{chen2020spatial} and SYSU-CD~\cite{shi2021deeply} datasets. The best two results are highlighted in \textcolor{red}{\textbf{red}} and \textcolor{blue}{\textbf{blue}}. All results are described as percentages (\%).}
        \label{tab:comp}
        \renewcommand\arraystretch{1.2}
        \resizebox{\linewidth}{!}{
        \begin{tabular}{r|c|cccc|cccc|cccc} 
        \hline
        \multirow{2}{*}{\textbf{Method~~~~~~~~~} } &\multirow{2}{*}{\textbf{Backbone}} &\multicolumn{4}{c}{ \textbf{WHU-CD }} &\multicolumn{4}{|c}{ \textbf{LEVIR-CD }} &\multicolumn{4}{|c}{\textbf{SYSU-CD}} \\ 
        & & $\textbf{F1}$ & $\textbf{P}$ & $\textbf{R}$ & $\textbf{IoU}$ & $\textbf{F1}$ & $\textbf{P}$ & $\textbf{R}$ & $\textbf{IoU}$ & $\textbf{F1}$ & $\textbf{P}$ & $\textbf{R}$ & $\textbf{IoU}$
  \\ 
        \hline
        FC-EF$_{18}$ \cite{daudt2018fully} &VGG &72.01 &77.69 & 67.10 & 56.26  & 83.4 & 86.91 & 80.17 & 71.53  & 75.07& 74.32& 75.84& 60.09\\ 
        FC-Siam-diff$_{18}$  \cite{daudt2018fully} &VGG & 58.81 & 47.33 & 77.66 & 41.66  & 86.31 & 89.53 & 83.31 & 75.92  & 72.57& 89.13& 61.21& 56.96\\ 
        FC-Siam-conc$_{18}$\cite{daudt2018fully} &VGG & 66.63 & 60.88 & 73.58 & 49.95  & 83.69 & 91.99 & 76.77 & 71.96  & 76.35& 82.54& 71.03& 61.75\\ 
        \hline
        IFNet$_{20}$\cite{zhang2020deeply}&VGG & 83.11 & 92.24 & 75.78 & 71.12  & 88.13 & 91.78 & 82.93 & 78.77  & 76.38& 82.44&72.38 & 61.85\\  
        BIT$_{22}$\cite{chen2021remote} &ResNet18 & 83.98 & 86.64 & 81.48 & 72.39  & 89.31 & 89.24 & 89.37 & 80.68  & 79.40& 82.32& 76.68& 65.84\\ 
        SNUNet $_{22}$\cite{fang2021snunet} &UNet++ & 83.50 & 85.60 & 81.49 & 71.67  & 88.16 & 89.18 & 87.17 & 78.83  & 79.96& 81.95& 78.08& 66.62\\ 
        ChangeFormer$_{22}$\cite{bandara2022transformer} &ViT & 89.88& 91.83& 88.02& 81.63& 90.40 & 92.05 & 88.80 & 82.48  & 77.83& 81.30& 74.65& 63.71\\
        TinyCD$_{23}$\cite{codegoni2023tinycd}&EfficientNet & 91.48& 92.22& 90.74& 84.30& 91.05& 92.68& 89.47& 83.57& 80.96& \textcolor{blue}{\textbf{85.71}}& 76.71& 68.01\\ 
        DGANet$_{23}$\cite{zhang2023difference}&CNN & 90.87& 95.59& 86.59& 83.27& 90.26& 92.03& 88.56& 82.25& 79.79& 78.86& \textcolor{blue}{\textbf{80.74}}& 66.37\\ 
        HSSENet$_{23}$\cite{yan2023hybrid}&ResNet18 & 91.55&94.56& 88.73&84.42&  91.48& 92.63& 90.35& \textcolor{blue}{\textbf{84.29}}&-&-&-&-\\
        WNet$_{23}$\cite{tang2023wnet}&ResNet18 & 91.25&92.37& 90.15&83.91&  90.67& 91.16& 90.18& 82.93&80.64&81.71&79.58&67.55\\
        TransY-Net$_{23}$\cite{yan2023transy}&Swin Transformer & 93.38&94.68&92.12&87.58&  \textcolor{blue}{\textbf{91.90}}& \textcolor{blue}{\textbf{92.90}}& 89.35& 83.64&\textcolor{blue}{\textbf{82.84}}&\textcolor{red}{\textbf{89.09}}&77.42&\textcolor{blue}{\textbf{70.71}}\\
        \hline
        \multirow{2}{*}{\textbf{RCTNet (Ours)}} &ResNet18 &\textcolor{blue}{\textbf{94.11}} &\textcolor{blue}{\textbf{95.84}} &\textcolor{blue}{\textbf{92.43}} &\textcolor{blue}{\textbf{88.87}} &91.33 &92.75 &\textcolor{blue}{\textbf{89.95}} &84.04 &82.23 &85.62 &79.10 & 69.83 \\
        &RegNet & \textcolor{red}{\textbf{95.05}}&\textcolor{red}{\textbf{96.04}}& \textcolor{red}{\textbf{94.09}}& \textcolor{red}{\textbf{90.57}}& \textcolor{red}{\textbf{91.93}}& \textcolor{red}{\textbf{92.91}}& \textcolor{red}{\textbf{90.97}}& \textcolor{red}{\textbf{85.07}}&\textcolor{red}{\textbf{83.01}} &84.33 & \textcolor{red}{\textbf{81.73}}& \textcolor{red}{\textbf{70.96}} \\
        \hline
        \end{tabular}}
    \end{center}
\end{table*}

\subsection{Paying Attention to Multi-Scale Features} \label{sec:pa_msf}
\noindent\textbf{Multi-Scale Fusion.}
Seeking to enhance the capture of temporal change information, we introduce the Multi-Scale Fusion (MSF) module, leveraging a series of convolution operations. As depicted in Fig.~\ref{img:U}~(a), MSF branches into four paths, each applying a convolution to the previous base feature map. These branches extract features at distinct scales, thereby enriching the final representation via concatenation. 
\begin{eqnarray}
\begin{aligned}
{c}' & =\operatorname{Conv}_{3 \times 3}\left(c_{in}\right), {c}'' =\operatorname{Conv}_{3 \times 3}\left({c}'\right),\\
{c}''' & =\operatorname{Conv}_{3 \times 3}\left({c}''\right), {c}'''' =\operatorname{Conv}_{3 \times 3}\left({c}''\right),\\
c_{out} & =c_{in}\oplus\operatorname{Cat}\left({c}',{c}'',{c}''',{c}''''\right)).
\end{aligned}
\end{eqnarray}
Further, we incorporate residual learning to preserve salient information and yield more expressive image features.

\noindent\textbf{Efficient Self-deciphering Attention.}
While the U-shape network architecture effectively combines multi-level features, its output lacks global semantic context, limiting CD accuracy. Inspired by SegFormer~\cite{xie2021segformer}, we introduce the Efficient Self-deciphering Attention (ESA) module to integrate global semantic relationships across decoder layers. Fig.~\ref{img:U} (b) illustrates ESA, which resembles a standard self-attention structure but adopts sequence reduction for computational efficiency. Assuming each of the heads $Q, K, V$ have the same dimensions $N \times C$, $N$ vectors each of dimension $C$, the self-attention is represented as:
\begin{eqnarray}
\begin{aligned}
    \operatorname{Attention}(Q, K, V)=\operatorname{Softmax}\left(\frac{Q K^{\top}}{\sqrt{d_{\text {head }}}}\right) V.
\end{aligned}
\end{eqnarray}

Traditionally, self-attention exhibits $\mathcal{O}\left(N^2\right)$ time complexity, where $N$ denotes the sequence length. To address this, ESA leverages sequence reduction with a reduction ratio $R$. This involves two steps: firstly, reshaping the key matrix $K$ from $N \times C$ to $\frac{N}{R} \times (C \cdot R)$ while preserving information. Secondly, a linear layer projects the reshaped matrix back to its original dimensions $\frac{N}{R} \times C$. Similar operations are applied to the value matrix $V$. Consequently, the time complexity reduces to $\mathcal{O}\left(\frac{N^2}{R}\right)$, significantly improving efficiency. In our experiments, we set $R$ to 4 for all decoder stages.

\subsection{Loss Function}
This paper utilizes a hybrid loss function to enhance training by combining the strengths of binary cross-entropy (BCE) and Dice losses. Our loss function is defined as:
\begin{eqnarray}
\begin{aligned}
\mathcal{L}=\mathcal{L}_{\text {bce}}+\mathcal{L}_{\text {dice}}.
\end{aligned}
\end{eqnarray}
The $\mathcal{L}_{\text{bce}}$ focuses on pixel-level classification accuracy and is formulated as:
\begin{eqnarray}
\begin{aligned}
\mathcal{L}_{\text {bce}}(p,g)=g\cdot \log \left(p\right)+\left(1-g\right) \cdot\log \left(1-p\right),
\end{aligned}
\end{eqnarray}
where $p$ is the predicted change map, $g$ is the corresponding ground truth, and $\cdot$ denotes element-wise product. The Dice loss emphasizes spatial overlap between predictions and ground truth, calculated as:
\begin{eqnarray}
\begin{aligned}
\mathcal{L}_{\mathrm{dice}}(p, g)=1-\frac{2 \cdot p \cdot g}{\left\|p\right\|_1+\left\|g\right\|_1},
\end{aligned}
\end{eqnarray}
where $\left\|\cdot\right\|_1$ denotes the $\ell_1$ norm. For our multi-stage predictions, the total loss is computed by summing the individual losses across all stages:
\begin{eqnarray}
\begin{aligned}
\mathcal{L}(p, g)=\sum_{i=1}^4 \mathcal{L}_{\mathrm{bce}}\left(p_i,g\right)+\sum_{i=1}^4\mathcal{L}_{\mathrm{dice}}\left(p_i,g\right),
\end{aligned}
\end{eqnarray}
where $p_i$ represents the predicted change map for stage $i$.

\begin{figure*}[t]
\centering
\includegraphics[width=1.0\textwidth]{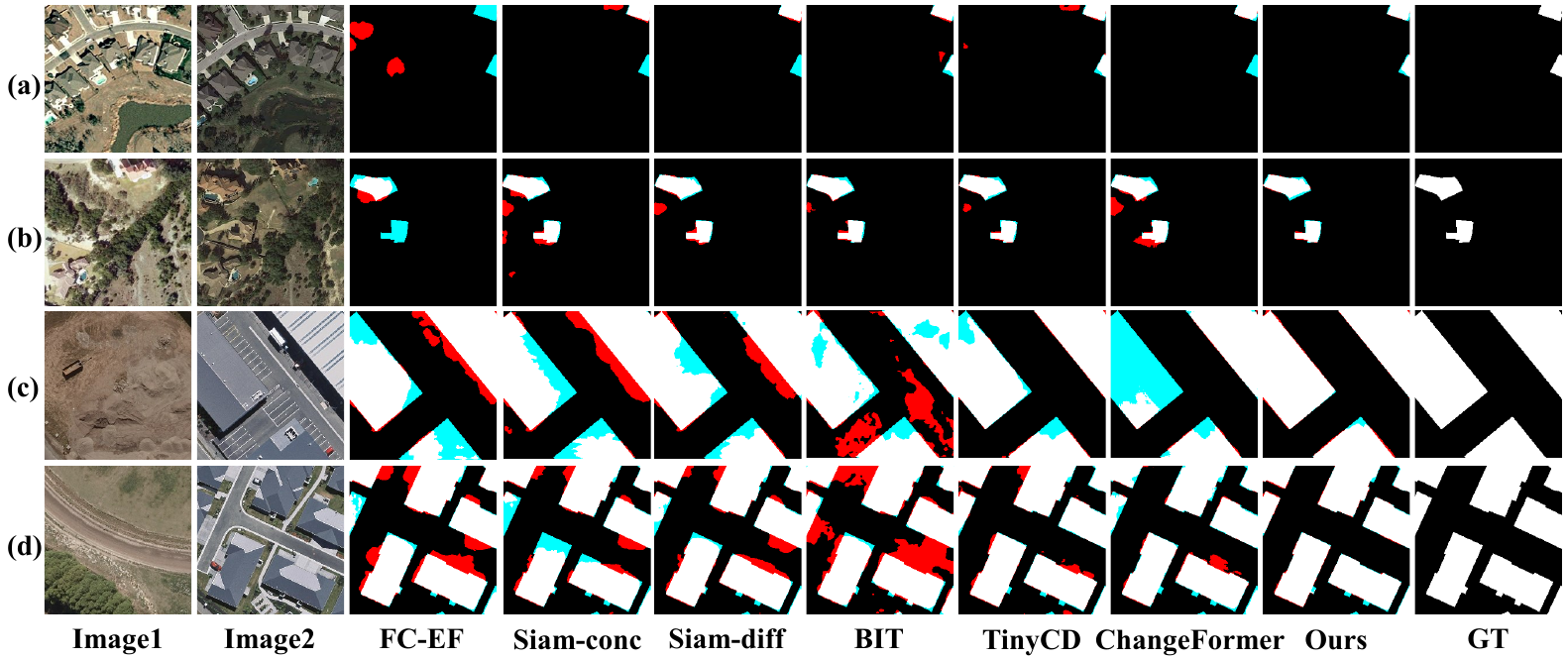}
\caption{Qualitative comparisons of our proposed method and state-of-the-art approaches on two benchmark datasets, LEVIR-CD~\cite{chen2020spatial} and WHU-CD~\cite{ji2018fully}. Examples (a) and (b) showcase results on LEVIR-CD, while (c) and (d) focus on WHU-CD. In each sample, ``GT'' represents the ground truth, white areas denote true positives, black areas represent true negatives, red areas indicate false positives, and blue areas represent false negatives.}
\label{img:comp}
\end{figure*}

\section{Experiments}
\subsection{Experimental Setup}

\noindent\textbf{Datasets.}
In this work, we evaluate our RCTNet on three challenging change detection benchmarks: WHU-CD~\cite{ji2018fully}, LEVIR-CD~\cite{chen2020spatial}, and SYSU-CD~\cite{shi2021deeply}. LEVIR-CD comprises 637 image pairs at 1024 × 1024 resolution, WHU-CD consists of a single image pair at
a larger size of 32 507 × 15 354, and SYSU-CD provides a substantial collection of 20,000 image patches ($256 \times 256$, 0.5m resolution) for CD tasks.

\begin{table}[t]
\begin{center}
\caption{Comparison results of the number of parameters (Params, M) and computational cost (FLOPs, G). The best result is highlighted in \textcolor{red}{\textbf{red}}.}
\label{tab:eff_comp}
\setlength{\tabcolsep}{1.5pt}
\renewcommand\arraystretch{1.3}
\resizebox{\linewidth}{!}{
\begin{tabular}{r|c|c|c|c|c} 
        \hline
        \textbf{Method~~~~~~~~~~~} & \textbf{Params} & \textbf{FLOPs} & \textbf{WHU-CD}&\textbf{LEVIR-CD}&\textbf{SYSU-CD} \\
        \hline
        IFNet$_{20}$\cite{zhang2020deeply} &50.71 &82.26& 83.11 & 88.13 & 76.38 \\  
        SNUNet $_{22}$\cite{fang2021snunet} &\textcolor{red}{\textbf{12.03}} &42.96& 83.50 & 88.16 & 79.96 \\ 
        ChangeFormer$_{22}$\cite{bandara2022transformer} &41.02 &202.87& 89.88 & 90.40 & 77.83 \\ 
        DGANet$_{23}$\cite{zhang2023difference} &12.28 &12.56& 90.87 & 90.26 & 79.79 \\ 
        WNet$_{23}$\cite{tang2023wnet} &43.07 &19.20& 91.25 & 90.67 & 80.64 \\ 
        \hline
        \textbf{RCTNet (ResNet18)} &14.6 &8.43& 94.11 & 91.33 & 82.23 \\ 
        \textbf{RCTNet (RegNet)} &13.8 &\textcolor{red}{\textbf{7.65}}& \textcolor{red}{\textbf{95.05}} & \textcolor{red}{\textbf{91.93}} & \textcolor{red}{\textbf{83.01}} \\ 
        \hline
\end{tabular}}
\end{center}
\end{table}
\noindent\textbf{Evaluation Metrics.}
To comprehensively evaluate the performance of our approach, we calculate four key metrics: Precision (P), Recall (R), F1-score (F1), and Intersection over Union (IoU). These metrics are defined as follows:
\begin{eqnarray}
\begin{aligned}
\rm{P}&=\frac{\rm{TP}}{\rm{TP}+\rm{FP}},~~~~~~\rm{R}=\frac{\rm{TP}}{\rm{TP}+\rm{FN}}, \\
\rm{F1}&=\frac{2\times \rm{P} \times \rm{R}}{\rm{P}+\rm{R}}, ~\rm{IoU}=\frac{T P}{F N+F P+T P},
\end{aligned}
\end{eqnarray}
where TP, TN, FP, and FN represent true positives, true negatives, false positives, and false negatives, respectively.

\noindent\textbf{Implementation Details.}
Our RCTNet implementation uses the PyTorch libraries~\cite{paszke2019pytorch} and utilizes an NVIDIA V100 GPU for training and testing. The network backbone is initialized with pre-trained weights from the RegNetY-1.6GF model~\cite{radosavovic2020designing} trained on ImageNet. We employ the Adam optimizer~\cite{kingma2014adam} with a momentum of 0.9, weight decay of 0.0001, and parameters $\beta _{1}$ and $\beta _{2}$ set to 0.9 and 0.99, respectively. The initial learning rate is 0.0005 and is dynamically adjusted during training using a poly learning rate decay with a power of 0.9. We train with a batch size of 32 for 50,000 iterations. Following previous works~\cite{daudt2018fully,fang2021snunet,bandara2022transformer,codegoni2023tinycd}, we divide the images in these datasets into patches of size $256 \times 256$. We apply random flipping, cropping, and temporal exchange to augment the training data to the input images.

\subsection{State-of-the-art Comparisons}
To comprehensively evaluate the effectiveness and efficiency of our proposed RCTNet for the change detection task on bitemporal remote sensing images, we benchmark it against several state-of-the-art approaches~\cite{daudt2018fully,zhang2020deeply,chen2021remote,fang2021snunet,bandara2022transformer,codegoni2023tinycd,zhang2023difference,yan2023hybrid,tang2023wnet,yan2023transy}.

\noindent\textbf{Quantitative Results.}
Compared to several state-of-the-art change detection approaches, our RCTNet surpasses them on benchmark datasets, as evidenced by Table~\ref{tab:comp}. 
RCTNet performs best in F1 and IoU metrics compared to existing methods.
In particular, for the WHU-CD dataset, our RCTNet (\textit{w/} RegNet) improves by \textbf{1.67} and \textbf{2.99} in terms of F1 and IoU metrics.
Furthermore, as depicted in Table~\ref{tab:eff_comp}, RCTNet showcases competitive computational efficiency while maintaining strong performance, featuring less number of parameters (Params) and floating point operations (FLOPs).

\noindent\textbf{Qualitative Results.}
Fig.~\ref{img:comp} presents visual comparisons of our method and other approaches on the LEVIR-CD and WHU-CD datasets. The presented approach skillfully retains the integrity of change information while preserving edge details through the effective extraction of the large-area and small-area change
information. Compared to previous methods, the proposed method demonstrates the capability to achieve relatively comprehensive detection results.

\subsection{Ablation Studies}
In this section, we conduct a series of ablation experiments on the LEVIR-CD validation set to scrutinize the effectiveness of the proposed components with RCTNet (\textit{w/} RegNet).
Table~\ref{tab:ablation} illustrates that the ablated versions of individual components exhibit varying degrees of performance degradation in comparison to the full model.
The ablation results demonstrate the effectiveness of our proposed modules.
Our introduction of RCTNet seamlessly merges features from various backbone network stages, enhancing semantic information and fine-grained details in multi-level feature maps.

\begin{table}[t]
    \centering
    \caption{Ablation studies of the proposed three key components on LEVIR-CD~\cite{chen2020spatial}. The best result is highlighted in \textcolor{red}{\textbf{red}}.}
    \label{tab:ablation}
    \setlength{\tabcolsep}{12.5pt}
    \renewcommand\arraystretch{1.3}
    \resizebox{\linewidth}{!}{
    {\fontsize{12}{14}\selectfont
    \begin{tabular}{c|c|cccc}
        \hline
        \# & \textbf{Component} & \textbf{F1} & \textbf{P} & \textbf{R} & \textbf{IoU } \\
        \hline
        1 &\textit{w/o} CSA     &91.52   &92.55   &90.50   &84.36 \\
        2 &\textit{w/o} MSF     &91.84   &92.78   &90.92   &84.91 \\
        3 &\textit{w/o} ESA     &91.67   &92.86   &90.52   &84.63 \\
        \hline  
        4 & \textbf{Full (Ours)} &\textcolor{red}{\textbf{91.93}}   &\textcolor{red}{\textbf{92.91}}   &\textcolor{red}{\textbf{90.97}}   &\textcolor{red}{\textbf{85.07}} \\
        \hline
    \end{tabular}}}
\end{table}

\section{Conclusion}
Our proposed deep learning approach, RCTNet, alleviates the limitations of existing change detection methods by mitigating inaccuracies from atmospheric variations, lighting changes, and phenological shifts. 
This impressive effort is achieved through a synergistic combination of the convolutional neural network and Transformer strengths.
Rigorous experimentation across diverse datasets demonstrates RCTNet's clear superiority over state-of-the-art models. It exhibits substantial accuracy improvements and a favorable trade-off between performance and computational cost.

\bibliographystyle{IEEEbib}
\bibliography{Main}

\end{document}